\documentclass[
    ,final            
  ]
  {aipproc}

\layoutstyle{6x9}

\usepackage{epsfig}
\begin{document}

\title{Summary of the COSY-11 Measurements\\ of Hyperon Production}


\author{D.~Grzonka \\ for the COSY-11 collaboration}{
  address={Institut f{\"u}r Kernphysik, Forschungszentrum J\"{u}lich, D-52425 J\"ulich, Germany}}

\begin{abstract}
The studies of hyperon production performed at COSY-11 are summarized.
The results of the 
experiments in the reaction channels
$pp  \rightarrow pK^+\Lambda$, $pp  \rightarrow pK^+\Sigma^0$, 
and $pp  \rightarrow nK^+\Sigma^+$ are shown.
Excitation functions from threshold up to about 90 MeV excess energies 
have been evaluated with high precision for the $\Lambda$ and $\Sigma^0$
production. The $\Lambda p$ and $\Sigma^0 p$ 
final state interactions were extracted.
The $\Sigma^+$ production was measured at 13 and 60 MeV excess energies.
\end{abstract}

\maketitle


\section{Introduction}
The hyperon-nucleon interaction is less known than the one for
the nucleon-nucleon system due to the difficulties in performing 
scattering experiments with the unstable hyperons.
The existing YN-scattering data are rather limited 
\cite{eng66,sec68,ale68,eis71}
and for a better understanding of the strong interaction in the 
nonperturbative region of the QCD an extension of the data base
in the strangeness sector is very important.
Besides hyperon-nucleon scattering, reactions into 3-body exit channels
like: $NN \rightarrow NKY$ can be used to extract detailed information
on the NY-subsystem.
The YN interaction is only one aspect covered by these kinds of experiments
which can be separated into three stages, the initial state interaction
of the incoming nucleons, the associated strangeness production process
and the final state interaction.
Final state interaction happens between all exit particles and by separating
a suitable kinematic region also the KN and KY interaction can be studied.
Furthermore information about the contributing reaction mechanisms
are obtained including the excitation of nucleon resonances which is
also directed to the structure of these resonances.
Most favourable for these studies are experiments close to the reaction
threshold due to the low relative momenta and therefore long interaction times
between the ejectiles. 
In order to get a detailed understanding of these elementary interactions
involving strangeness differential cross sections as a function
of spin and isospin degrees of freedom are required.
A significant contribution to this kind of physics 
has been done by the hyperon production
experiments at COSY-11.   

\section{The experimental setup for strangeness production at COSY-11}

\begin{figure}
  \includegraphics[width=.6\textwidth]{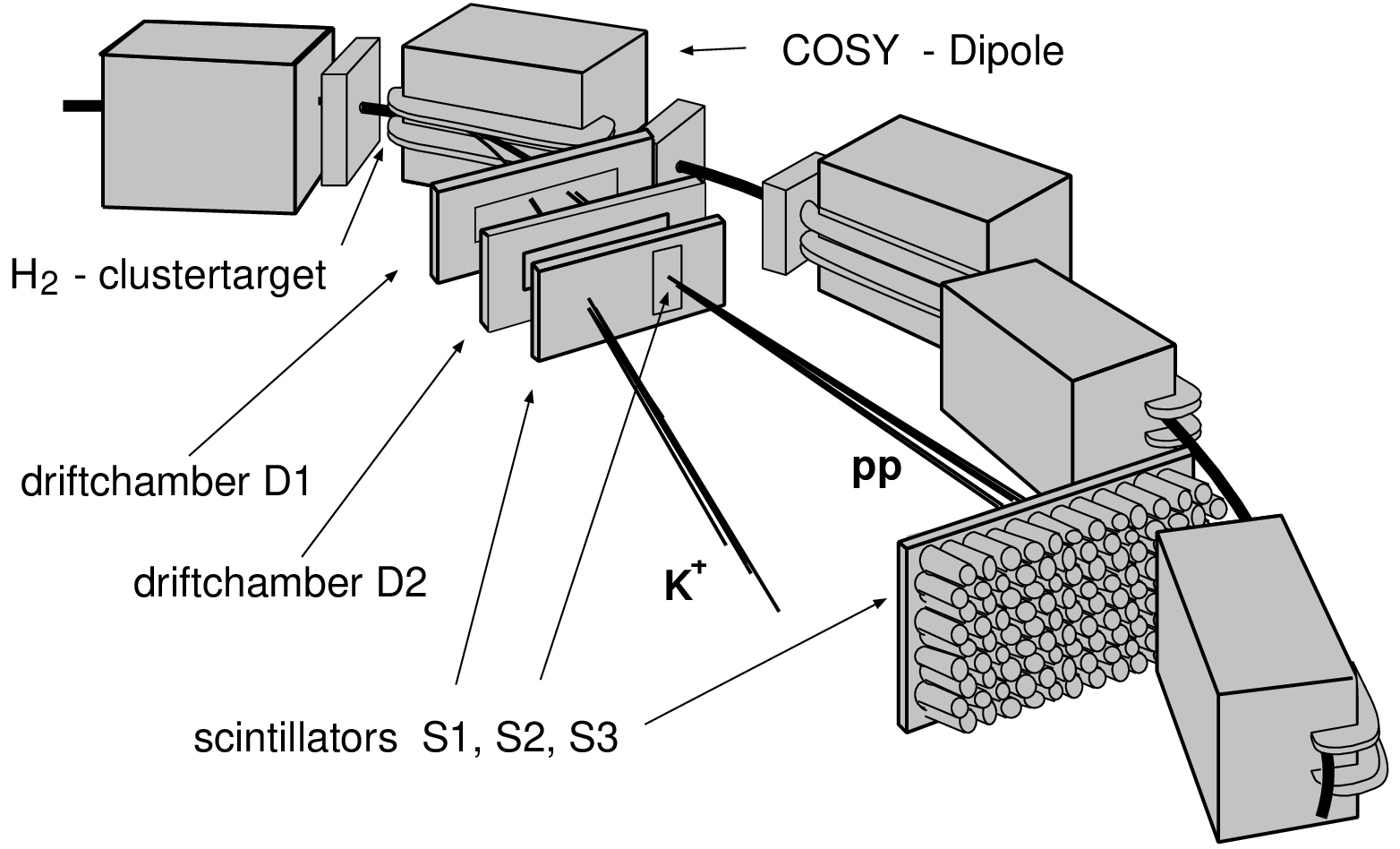}
  \includegraphics[width=.4\textwidth]{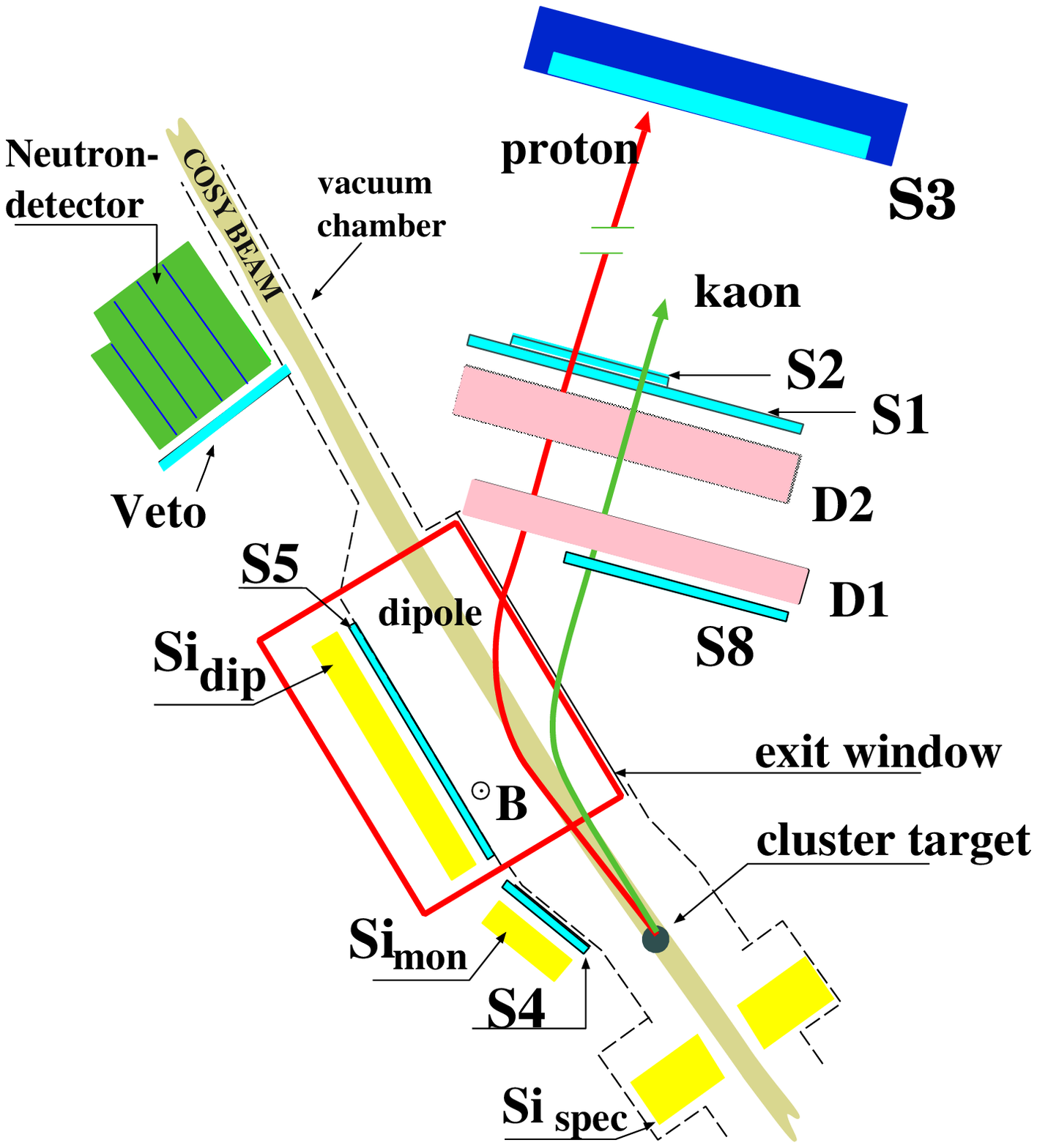}
  \caption{
\label{c11setup}
The COSY-11 detection system installed at a COSY machine dipole
with the detector components relevant for the hyperon production studies.
The left side shows a 3-d view of the arrangement and on the right side 
is a sketch of the detector components to illustrate the principle of
operation.
The S8  scintillator was only used for the $pp \rightarrow n K^+ \Sigma^+$ reaction.}
\end{figure}

The internal COSY-11 installation \cite{bra96} at COSY \cite{mai97}
was designed for near threshold meson 
production studies. It used a COSY machine dipole as magnetic spectrometer
and included scintillation detectors and drift 
chambers to reconstruct particle tracks of positively charged 
particles and measure their velocities in order to determine their 
four-momentum components with high precision. A sketch of the setup
is given in fig. \ref{c11setup} and for more details see \cite{bra96}.

\begin{figure}
\begin{minipage}[c]{0.55\textwidth}
  \includegraphics[width=\textwidth]{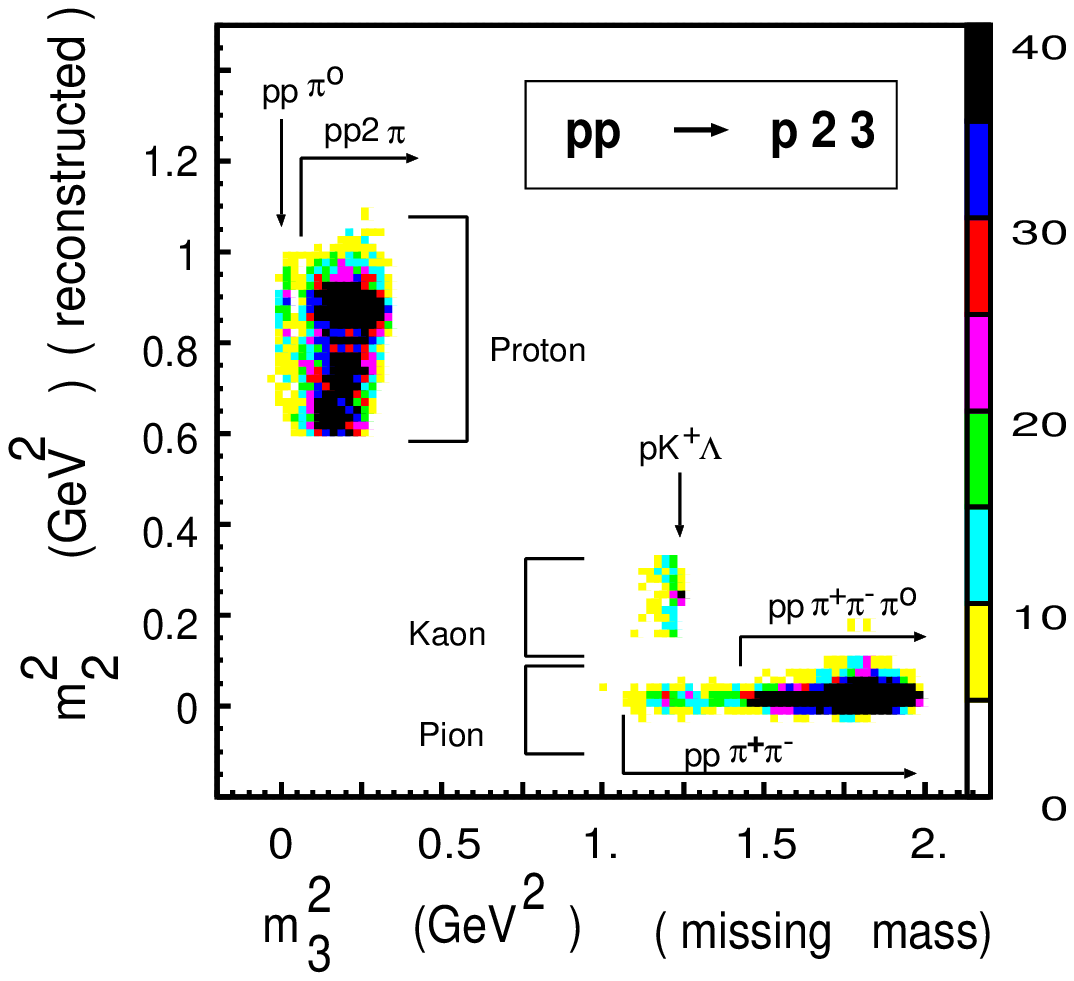}
\end{minipage}
\begin{minipage}[c]{0.45\textwidth}
\centering 
\includegraphics[width=\textwidth]{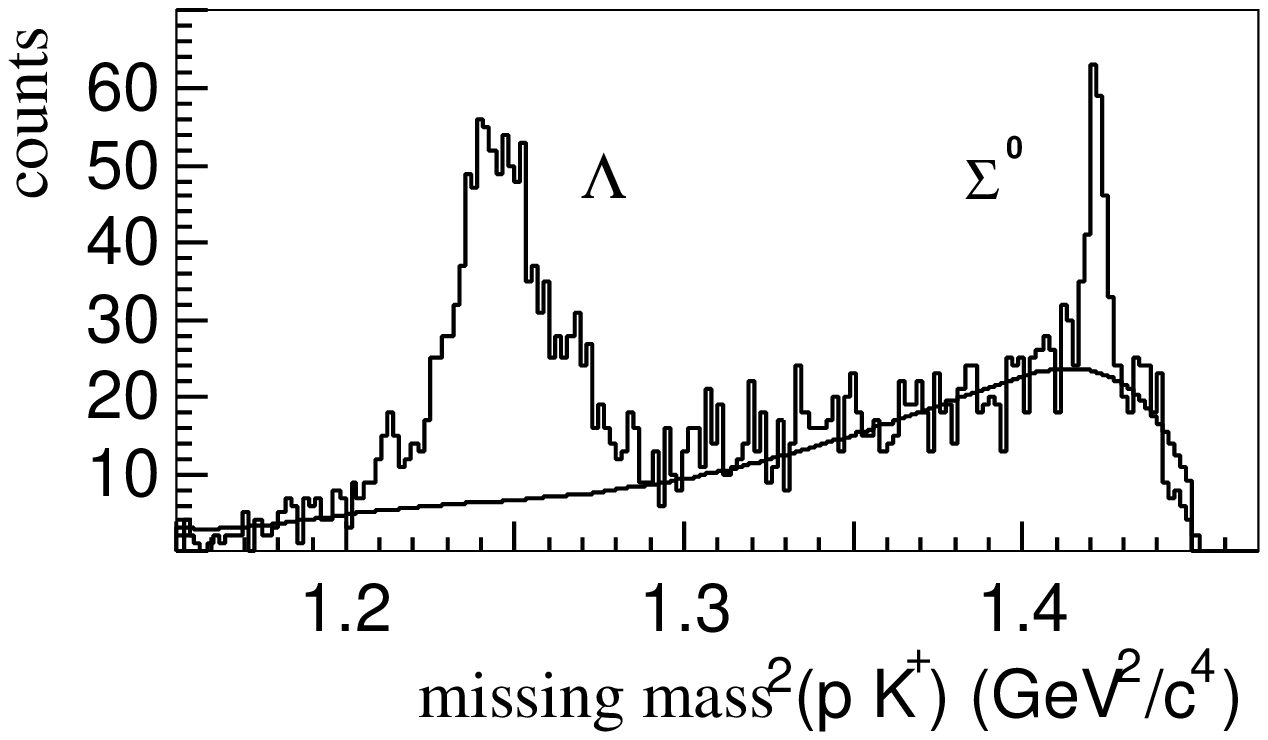}
\includegraphics[width=\textwidth]{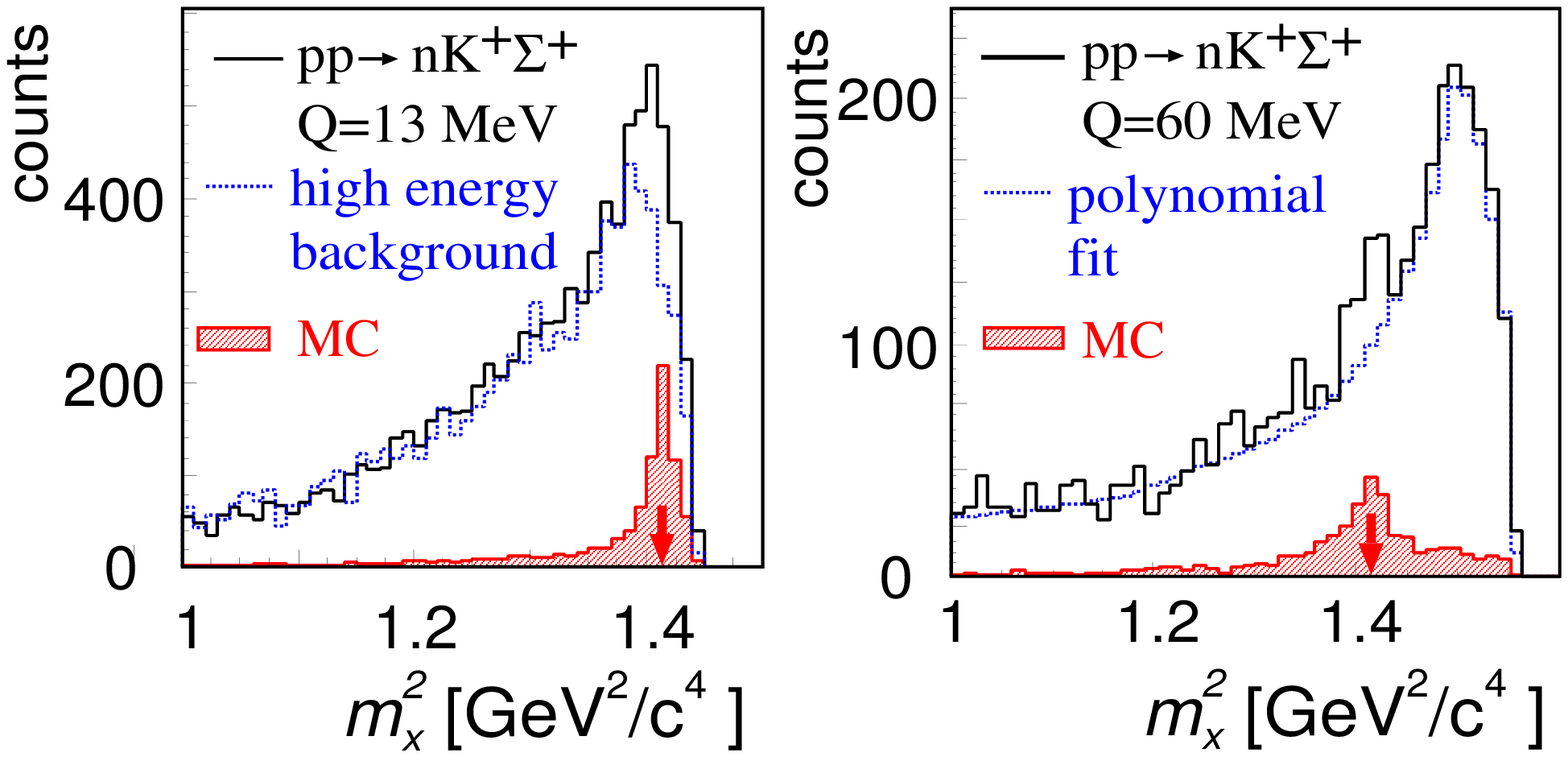}
\end{minipage}
  \caption{\label{hypexp}Invariant mass of the second track as a function of the
missing mass for $\Lambda$ production at 7 MeV excess energy (left). 
Missing mass squared distribution for $\Sigma^0$ production
at 7 MeV excess energy (up right).
Missing mass squared distributions for $\Sigma^+$ production
at 13 and 60 MeV excess energy (down right) with the applied 
background subtraction and the expected distributions of 
$\Sigma^+$ production from 
Monte Carlo.}
\end{figure}

In the case of hyperon production via the reaction channels 
$pp \rightarrow pK^+ \Lambda / \Sigma^0$ 
the proton velocities are measured with the scintillator 
hodoscopes S1 and S3 but for the kaon the $\sim$~9 m flight path to S3 is 
too long. Most of the kaons would decay before reaching S3. Here the 
flight path from the target to S1 is used where the start time is 
calculated from the measured proton momentum.
The particle identification is worse than in the proton case 
due to the much shorter flight path but its still sufficient 
to separate most of the pions and protons from the kaons.
The hyperon four-momentum $P_{hyperon}$ is determined 
by the missing mass technique:
$P_{\Lambda} = P_{beam} - P_{p} - P_{K^+}$
with the known beam $P_{beam}$ and the measured proton $P_{p}$ 
and kaon $P_{K^+}$ four-momenta.
This method results in a rather clean separation
of the hyperon production events as can 
be seen from fig. \ref{hypexp} left for an event sample of $\Lambda$ production
at 7 MeV excess energy.
In the case of the $\Sigma^0$ production its similar but the background level is
higher as can seen from fig. \ref{hypexp} up right which shows the missing mass
squared distribution 
in the kaon band at an excess energy of 7 MeV for $pp \rightarrow pK^+ \Sigma^0$.
In parallel to the $\Sigma^0$ production also $\Lambda$ production
at about 80 MeV higher excess energies is measured.

With the addition of a neutron detector, installed for studies 
at a deuteron target, another hyperon channel was accessible
at COSY-11, namely the $pp  \rightarrow nK^+ \Sigma^+$ reaction.
Here the peak to background ratio was less favourable,
see fig. \ref{hypexp} down right,
because no proton is in the exit 
channel to produce a precise timing signal.
The neutron detector provided the 
time and position of the point of the first neutron interaction 
producing a charged ejectile from which the neutron momentum was calculated.
The absolute time calibration was performed with $\gamma$'s by selecting 
$pp  \rightarrow pp \pi^0$ events with the $\pi^0$ decaying 
within the target into two $\gamma$'s from which the event start time
was calculated. For the $K^+$ time of flight measurement
the additional S8 scintillator was used with a distance of only 1.9 m to S1.
The $\Sigma^+$ with a $c\tau$ of 2.4 cm couldn't be measured directly
but its four-momentum was determined by a missing mass analysis.

Further hyperon channels are not feasible at COSY-11. In principle also the 
($n \Lambda$) and ($n \Sigma^0$) system could be studied by using 
a deuteron target
but the additional detection of the spectator proton would
reduce the efficiency drastically.
Also hyperon decay products could in principle be measured but 
the efficiency was extremely low.

In all measurements the luminosity was determined by elastic $pp$-scattering
detected in parallel to the hyperon production. For the detection of the
second proton a Si-pad detector combined with a scintillator
($Si_{mon} / S4$ in fig. \ref{c11setup} ) was installed.
For studies with a polarized beam the polarization has to be 
determined. Two detection systems served for this aim:
In addition to the COSY polarimeter a pair of wire chambers
and scintillators were installed above and below the beam close to
the target to measure the elastic $pp$-scattering at $\phi = 90 ^{\circ}$
which is independent of the polarisation.

\section{Experimental results}

When COSY-11 went into operation in 1996 no data were available for 
$\Lambda$ and $\Sigma$ hyperon production close to the reaction threshold.
For the reaction channel $pp \rightarrow pK^+ \Lambda$ 
above 300 MeV/c excess energy 
data were existing mostly from bubble 
chamber measurements at CERN \cite{bal88}.
On the theoretical side 
parametrizations of the excitation function 
were on the market which differ close to threshold by several 
orders of magnitude \cite{ran80,sch88}.
For the $\Sigma$ production the situation was similar, there were no data 
available below a few hundred MeV excess energy.

\begin{figure}
  \includegraphics[width=.7\textwidth]{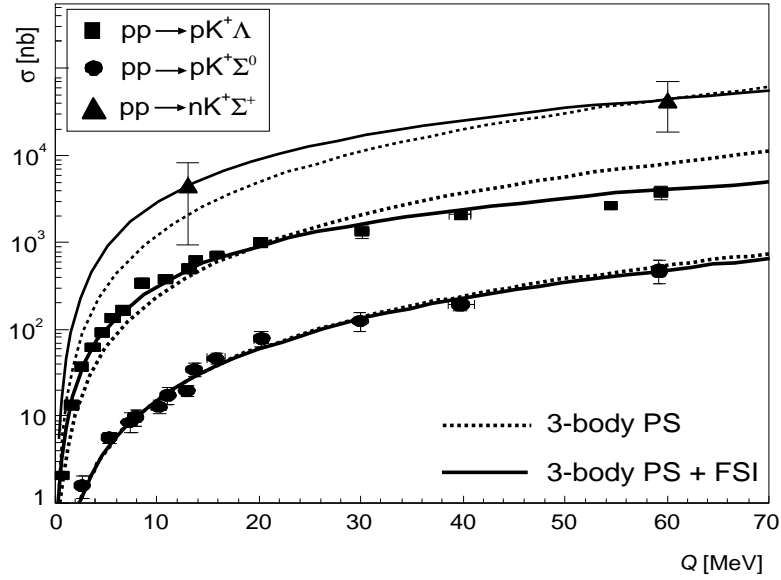}
\caption{
\label{c11data}
The $pp \rightarrow pK^+ \Lambda$, $pp \rightarrow pK^+ \Sigma^0$ 
and $pp \rightarrow nK^+ \Sigma^+$ cross
sections as a function of the excess energy $Q$ 
~\cite{bal98,sew99,kow04,bil98}.
The  lines show the calculations corresponding to 3-body phase space with
(solid line) and without (dashed line) final state interaction. }
\end{figure}

The first hyperon production studies at COSY-11 were performed for the
$\Lambda$ channel. The excitation function for the 
$pp \rightarrow pK^+ \Lambda$ reaction was measured in several beam times 
for excess energies between 0.7 MeV and 90 MeV 
\cite{bal96,bal98,sew99,kow04}. 
Compared to the parametrization of \cite{ran80,sch88}
the data differ by more than an order of magnitude but
with the COSY activities in the hyperon channel several theory
groups were triggered to develop improved models which describe the data 
much better \cite{sib95,lik95,lik98, 
fal97,tsu97,tsu99}.
In fig. \ref{c11data} the threshold data are shown in a 
linear excess energy scale
including the expected phase space behavior with
and without $NY$-FSI adjusted to the data.
It is clearly seen that close to threshold
a pure 3-body phase space description is insufficient for the 
case of the $\Lambda$ hyperon production.
The final state interaction between proton and $\Lambda$ has to be taken
into account. To include FSI 
the F\"aldt-Wilkin parametrization has been used \cite{fal97,wil07}.

\begin{figure}
  \includegraphics[angle=-90,width=.8\textwidth]{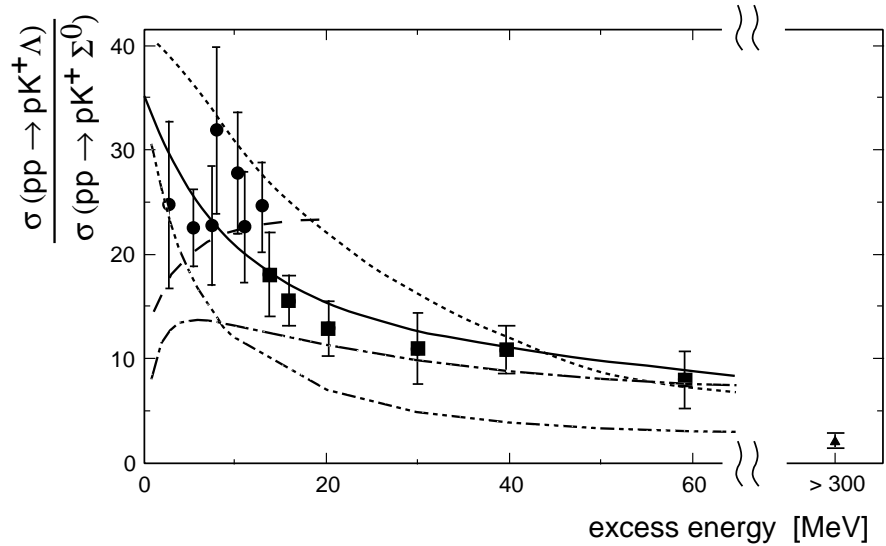}
\caption{
\label{lsratio}
Cross section ratio for $\Sigma^0$ and $\Lambda$ production in the threshold
region compared to different model predictions.
The data are from COSY-11 (solid circles \cite{sew99}, solid squares
\cite{kow04}) and a mean value from \cite{bal88} (solid triangle).
The curves represent calculations within different models,
$\pi$ and $K$ exchange with destructive interference
~\cite{gas01} (dashed line),
incoherent $\pi$ and $K$ exchange~\cite{sib99} (dashed-dotted),
meson exchange with intermediate $N^*$ excitation 
~\cite{sib99,tsu99}
(dashed-double-dotted) and
effective Lagrangian approach including $N^*$ excitation~\cite{shy01,shy06}
(dotted line). The solid line is the ratio of the phase space behavior
including $\Lambda p$ FSI.
}
\end{figure}

Similar data were taken for the $\Sigma^0$ production channel
$pp \rightarrow p K^+ \Sigma^0$ \cite{sew99,kow04}.
In a first study
for excess energies around 15 MeV the obtained results (fig. \ref{lsratio})
show the remarkable feature that the cross section
ratio  $\sigma _{\Lambda} / \sigma_ {\Sigma^0}$
gives a factor of 28 instead of $\sim$~2.5 known from high energy
data above 300 MeV excess energy expected from isospin relations.
The first step to understand this behavior was the extension to
higher excess energies to see the transition from this unexpected
high ratio to the value of 2.5.

In order to reduce systematical errors 
further data were taken in
the supercycle mode of COSY 
operation which allows to switch the beam momentum betweeen two 
successive COSY cycles.
One cycle for $\Lambda$ production was followed by a few cycles for $\Sigma^0$ 
production at the same excess energy.
The data are also given in fig.\ref{c11data}.

In the $\Sigma^0$ case there is no need for any FSI, an inclusion of 
(p$\Sigma$) FSI doesn't give any improvement of the $\chi ^2$ in the fit.
The difference between the (p$\Sigma$) and (p$\Lambda$) system is best seen
in the cross section ratio which is shown in fig. \ref{lsratio}.
The cross section ratio
$\sigma _{\Lambda} / \sigma_ {\Sigma^0}$
goes smoothly down to the low energy value.

The question arises whether the whole enhancement in the $\Lambda$ channel
is due to the strong (p$\Lambda$) FSI 
as proposed by \cite{sib06}
or the production process
itself gives some enhancement for the $\Lambda$ channel.
If its a pure FSI effect why is the (p$\Lambda$) so much larger
than the (p$\Sigma$) FSI?
When the first data on the cross section ratio were published
several theory groups tried to describe the data in different
models:
coherent or incoherent $\pi$ and $K$
exchange or in addition an intermediate resonance excitation.
Most of them indeed show a trend of increasing ratio towards the threshold,
see fig. \ref{lsratio},
but there is no clear preference of any description suggested.

More data are needed to understand the hyperon production process
as for instance data in other isospin channels.
To illustrate this lets consider a specific model. In the
J\"ulich model \cite{gas01} the $pp \rightarrow p K^+ \Lambda / \Sigma^0$ reactions are
described by only pion and kaon exchange where a reduction of the $\Sigma^0$ cross
section results from a destructive interference of the pion and kaon amplitudes.
Calculations for the $pp \rightarrow n K^+ \Sigma^+$ channel within this
model show a big difference between a destructive 
($\sigma_{pp\rightarrow nK^+ \Sigma^+} / \sigma_{pp\rightarrow pK^+ \Sigma^0}
\ = 3.1$) and a constructive
($\sigma_{pp\rightarrow nK^+ \Sigma^+} / \sigma_{pp\rightarrow pK^+ \Sigma^0}
\ = 0.34$) interference.
A similar high sensitivity is expected 
also in other models which include nucleon resonances.

Therefore the study was extended to the $\Sigma^+$ production in order 
to disentangle the different production mechanisms \cite{roz06}.
From the experimental point of view a clean separation of the
 $pp \rightarrow n K^+ \Sigma^+$ channel
was difficult due to the high background level which resulted
in large error bars for the cross section determination.
However, in spite of the large errors the measured cross sections 
exceed the highest predictions from models 
\cite{gas01,sib99,tsu99,sib07} by
at least an order of magnitude. 
The data points are shown in fig. \ref{c11data}.
This interesting but not yet understood result confirms the need
for more data in the hyperon sector.

As mentioned in the introduction the three body final state allows
to study within some approximation the individual two body subsystems.
This was done by a Dalitz plot analysis of the COSY-11 $\Lambda$ production data
in order to extract the $\Lambda$-p scattering length \cite{bal98b}.
Within such an analysis with unpolarized beam and target
it is not possible to separate the
spin singlet and spin triplet components of the scattering length.
Only spin averaged values for scattering length $\bar{a}$ and effective range
$\bar{r_0}$ could be determined with values of
$\bar{a}\, = \, -2.0 fm$ , $\bar{r_0} = 1.0 fm$.
The analysis was done in 
an effective range expansion which, however,
is only applicable for systems where the
scattering length is significantly larger than the effective range.
Furthermore  the procedure exhibits strong correlations between the
effective range parameters $a$ and $r$ that can only be disentangled by
including $\Lambda N$ elastic cross sections data.

A new method to determine the scattering length was
developed by the J\"ulich \mbox{theory} group which is based on a dispersion relation
technique \cite{gas04,gas05}. The advantage of the technique is that the error
in the relation between scattering length and experimental observable
due to the derivation in the model is well under control and is below
0.2 fm. Furthermore observables have been derived where spin singlet 
(($1+A_{xx} + A_{yy} + A_{zz}) \cdot \sigma _{\Theta (K) = 90 ^{\circ}}$)
and spin triplet 
($A_{0y} \cdot \sigma _{\Theta (K) = 90 ^{\circ}}$) 
contributions are separated.
The measurement of a separate spin singlet contribution requires all spin correlation
coefficients ($A_{ii}$), i.e. polarized beam and polarized target, but for the spin triplet
component the asymmetry $A_{0y}$ is sufficient if a special kinematic condition,
kaon emission around 90 deg. is selected.

In order to extract the spin triplet scattering length of
the $\Lambda p$ system a measurement with polarized beam
has been performed at COSY-11.
The experiment aimed for a precision of the scattering length
in the same order as the theoretical uncertainty of about 0.2 fm.
The error is given by the statistical  uncertainty 
and by the size of the asymmetry which is not known in this excess energy
region.

From first results of the analysis the asymmetry seems to be rather small which would result
in large errors on the scattering length but the 
data evaluation  
is still going on
and final results have to be awaiten.

Besides COSY-11 also other COSY experiments studied hyperon production.
The TOF collaboration uses a large acceptance non magnetic detection system 
with a decay spectrometer for the delayed strange particle decays
resulting in a high selectivity for the hyperon reaction channels.
Various reactions were studied like $pp \rightarrow p K^+ \Lambda$
\cite{bil98,abd06},
$pp \rightarrow p K^+ \Sigma^0$, $pp \rightarrow p K^0 \Sigma^+$ \cite{abd07},
and $pp \rightarrow n K^+ \Sigma^+$ but at somewhat higher excess energies
than COSY-11.
For the $pp \rightarrow p K^+ \Lambda$ reaction channel at
excess energies of 85, 115 and 171 MeV Dalitz plot analyses
have been performed which show clearly the excitation of nucleon
resonances contributing to the production process \cite{abd06}. 
But to what extent the resonance excitation contributs
to the COSY-11 data close to threshold is not clear.
At the BIG KARL magnetic spectrometer an inclusive measurement of
\mbox{$pp \rightarrow p K^+ \Lambda$} reaction was done by detecting the kaon
with high momentum resolution in order to determine the 
$p\Lambda$ scattering parameters. The analysis gives constrains on
the singlet and triplet scattering length where $|a_s| > |a_t|$ \cite{hin05}
but to separate singlet and triplet contribution measurements
with polarized beam and target are needed.
Furthermore at ANKE, an internal magnetic spectrometer, hyperon production
studies have started. The kaon detection is done by measuring the delayed
decay of stopped kaons which gives a very high selectivity for kaons.
Data of the $pp \rightarrow n K^+ \Sigma^+$ reaction have been taken at
93 and 128 MeV excess energy which don't show such a high cross section
as the COSY-11 data at lower excess energies but are consistent with
model predictions \cite{val072}. These studies will be continued 
at lower excess energies \cite{val07}.
\section{Conclusion}

The hyperon production studies of the COSY-11 collaboration 
resulted in precise cross section data of the reaction channels
$pp \rightarrow p K^+ \Lambda$ and  $pp \rightarrow p K^+ \Sigma^0$
from the production  threshold up to about 90 MeV excess energy
where no data were available at all.
These cross section data are important ingrediants for calculations
in different fields like heavy ion collision studies, hypernuclei production
or neutron star formation.

The excitation functions allowed to extract the $NY$ final state interaction
which is related to the $NY$ interaction strength parameterized by the
scattering length. Models of the hyperon-nucleon scattering 
mostly rely on SU(3) symmetry relations for the coupling constants and 
fit the model to the data.
There is need for an improved data base
in this strangeness sector for a better
understanding of the strong interaction in this
domain of non perturbative QCD.
 
For the $p\Lambda$ system a strong $p\Lambda$ FSI is clearly seen
whereas the $p\Sigma^0$ seems to show no FSI although the
quark content of both systems is the same.
But for a clear picture also the other $\Sigma$ isospin channels
have to be studied.  A first attempt has been done
with the measurement of the $pp \rightarrow n K^+ \Sigma^+$ reaction
cross sections which exceed largely the model predictions.
The large error bars indicate somehow the limit of the COSY-11
facility for these studies.
An improved detection technique has to be applied for further investigations.

Studies of the hyperon channels have to be continued. There are still many
open questions and more differential observables 
including the spin and isospin degrees of freedom 
are needed to disentangle the contributing production mechnisms
and to improve the knowledge on the $YN$-interaction.
At ANKE and at the TOF experiment \cite{tof07} 
further studies are proposed
and in the near future certainly also WASA at COSY will be 
used for such investigations. With its large solid angle coverage for charge 
and neutral particles in principle 
all hyperon channels can be studied in detail.

\section{Acknowledgments}

        \label{acknow}
The hyperon production studies at COSY-11 were supported by the 
following grants:
European Community - Access to 
Research Infrastructure action of the Improving \mbox{Human} Potential
Programme (FP5 and FP6 within HadronPhysics RII3-CT-2004-506078, 
FFE grants (41266606 and 41266654, 41324880) from the Research
Center J\"ulich, DAAD Exchange Programme (PPP-Polen),
Polish State Committee for Scientific Research 
(grant No. PB1060/P03/2004/26, 2P03B-047-13),
the German Research Foundation (DFG) grant No. GZ: 436 POL 113/117/0-1,
the International B{\"uro} IB-DLR \mbox{(PL-N-108-95)},
and the
Verbundforschung of the BMBF (06MS881I).
%




\end{document}